\documentclass[10pt]{article}

\usepackage{amsmath, amssymb, amsthm}
\usepackage{graphicx,color}
\usepackage{subfig}
\usepackage{tabularx}
\usepackage{url}
\usepackage[super,sectionbib,comma,sort&compress]{natbib}
\usepackage[top=3.0cm,bottom=2.5cm,left=3cm,right=3cm]{geometry}

\begin{document}

\title{A multi-resolution model to capture both global fluctuations of an enzyme and molecular recognition in the ligand-binding site}

\author{Aoife C.\ Fogarty$^{\dagger}$, Raffaello Potestio$^{\dagger}$, Kurt Kremer*$^{\dagger}$}
\date{}
\maketitle
$\dagger$ {\it Max Planck Institute for Polymer Research, Ackermannweg 10, 55128 Mainz, Germany}

* {\it Corresponding author}, kremer@mpip-mainz.mpg.de\\

\noindent This is the pre-peer reviewed version of the following article: \\

Fogarty, A. C., Potestio, R. and Kremer, K. (2016), A multi-resolution model to capture both global fluctuations of an enzyme and molecular recognition in the ligand-binding site. Proteins. doi: 10.1002/prot.25173, \\

\noindent which has been published in final form at \url{http://onlinelibrary.wiley.com/wol1/doi/10.1002/prot.25173/full}. This article may be used for non-commercial purposes in accordance with Wiley Terms and Conditions for Self-Archiving.\\

\abstract{In multi-resolution simulations, different system components are simultaneously modelled at different levels of resolution, these being smoothly coupled together. In the case of enzyme systems, computationally expensive atomistic detail is needed in the active site to capture the chemistry of substrate binding. Global properties of the rest of the protein also play an essential role, determining the structure and fluctuations of the binding site; however, these can be modelled on a coarser level. Similarly, in the most computationally efficient scheme only the solvent hydrating the active site requires atomistic detail. We present a methodology to couple atomistic and coarse-grained protein models, while solvating the atomistic part of the protein in atomistic water. This allows a free choice of which protein and solvent degrees of freedom to include atomistically, without loss of accuracy in the atomistic description.
This multi-resolution methodology can successfully model stable ligand binding, and we further confirm its validity via an exploration of system properties relevant to enzymatic function. In addition to a computational speedup, such an approach can allow the identification of the essential degrees of freedom playing a role in a given process, potentially yielding new insights into biomolecular function.}

\newpage

\section{Introduction}

Biomolecules in solution are complex, heterogeneous systems with length and timescales covering many orders of magnitude.\cite{Atilgan_BiophysJ_2009-protdyn_timescales,McCammon_AnnRevBiochem_1983-dyn_prot_fn,MacCallum_Science_2012-protfold} 
Simulating such systems therefore often leads to requirements that are difficult to reconcile, namely (i) large systems and long simulation times, and (ii) accurate models that contain sufficient physical and chemical detail to describe a given phenomenon, and that are therefore computationally expensive. As such, biomolecular simulations may benefit from a concurrent multi-resolution approach. This involves identifying those parts of the system where the physical and chemical detail plays an essential role in the phenomenon of interest, and describing them using a sufficiently high-resolution model, while using a less detailed, computationally more efficient model for the remainder of the system.\cite{vanGunsteren_AngewChimie_2013-multisc_bio_review,Tozzini_AccChemRes_2010-multiscale_prot} Conversely, instead of merely obtaining computational speedup, the goal may in fact be to identify precisely which degrees of freedom play a role in a biomolecular process, via a model in which the degrees of freedom included at a certain resolution can be arbitrarily varied.

In most concurrent multi-resolution simulation approaches, each distinct system component, such as protein, nucleic acid, lipid membrane, or aqueous solvent, is modelled in its entirety using one level of resolution, e.g. an atomistic protein in a coarse-grained solvent or embedded in a coarse-grained membrane,\cite{Schaefer_JPCB_2013-prot_aacg_electrost,vanGunsteren_EurBiophysJ_2012-cgwat_atomprot,Voth_JPCB_2006-cgaa_ionchannel} or a coarse-grained protein in a continuum environment.\cite{Atzberger_JCompPhys_2013-hybrid_continpart} However, when the goal is to construct a model which includes only the minimum possible number degrees of freedom, one must be able to place boundaries between resolutions at any arbitrary place within the system. Take for example the case of enzymatic function. In its simplest form an enzyme can be seen as being composed of two parts: an active site, at which the substrate binding and catalytic reaction occur, and the remainder of the enzyme, which exists to induce in the active site those properties necessary for its function; these may include its structure, conformational fluctuations, dynamics, or electric field.\cite{Warshel_Proteins_2010-dawn21st,Truhlar_science_2004-howenzwork}
Furthermore, the aqueous solvent is known to play an essential role in enzymatic function.\cite{Dordick_PNAS_1992-wat_flex_polar,Klibanov_TrendsBiotechnol_1997-enz_org_sol}
An accurate model of substrate binding therefore requires at minimum an atomistic level of detail in the description of the substrate, binding site and neighbouring water molecules. However, the global structure and conformational fluctuations of a protein can be captured on a more coarse-grained level,\cite{Tirion_PRL_1996-firstENM,Grossfield_Proteins_2011-enm_md} as can the aqueous solvent further away from the binding site.\cite{Kremer_JCP_2015-adresprot,vanGunsteren_JCTC_2015-flex_multiscale} 

Here we propose a new approach which allows one to model at an atomistic resolution only the precise subset of degrees of freedom really necessary for the study of a given phenomenon, even when this leads to a boundary between resolutions which falls within a macromolecule, or includes only part of the solvent. For the enzyme system studied here, this means a coarse-grained protein in which is embedded an atomistic binding site, solvated by a small sphere of atomistic water, which freely exchanges with the reservoir of coarse-grained water filling the remainder of the simulation box.
In the construction of such a multi-resolution model, two methodological issues arise: (i) the coupling between particles with different resolutions connected by bonds, and (ii) how to describe solvent particles diffusing between regions at different levels of resolution. The former issue has been explored in a small number of polymer\cite{Carbone_JChemPhys_2012-mixfixresolution,Kremer_PhysRevE_2003-dualres_bisphenol} and biomolecular\cite{Carloni_PhysRevLett_2005-cg_atom_prot,Carloni_BiophysJ_2008-dualres,Pantano_PCCP_2011-simultan_aacg_dna,Pantano_JCTC_2015-dualres} studies. 
In particular, Neri and co-workers developed a model in which an atomistically detailled active site was incorporated into a coarse-grained G\={o} model.\cite{Carloni_PhysRevLett_2005-cg_atom_prot} Here, we present a method for inserting atomistic residues into a protein whose essential structure and conformational fluctuations are described using an Elastic Network Model (ENM). We note that the approach presented here differs from that of Neri et al.\cite{Carloni_PhysRevLett_2005-cg_atom_prot} both in our treatment of the atomistic--coarse-grained coupling within the protein, and in our inclusion of explicit solvent.
The second methodological issue, that of allowing solvent particles to change their resolution on the fly, can be tackled using the Adaptive Resolution Scheme (AdResS),\cite{Kremer_JCP_2005-onthefly} in which interpolation of atomistic and coarse-grained forces across a transition region allows for a smooth coupling between different resolution levels, and free diffusion of solvent particles throughout the simulation box. Only the small subregion of water solvating the atomistic protein active site is then described at an atomistic level, but behaves as though it were in a fully atomistic system.\cite{Clementi_JPhysCondensMat_2007-adres_water,DelleSite_JCP_2008-diffusive_water,DelleSite_PhysRevX_2013-GCadres,Donadio_PRL_2013-hadres_molliq}

This multi-resolution approach not only leads to greater computational efficiency via both a reduction in the number of degrees of freedom simulated, and via the improved efficiency in the sampling of slow, large-scale protein fluctuations which is inherent in the use of a coarse-grained protein model. It can also allow the simulation of large biomolecular systems where atomistic structure is not known everywhere and where atomistic simulations are not even possible, but where low-resolution experimental data can still be exploited to parametrise coarse-grained models for some parts of the system. 

\section{Methods}
\label{sec: model}

We demonstrate our methodology on an aqueous solution of hen egg-white lysozyme (HEWL), a widely studied 14~kD enzyme that hydrolyses glycosidic bonds in polysaccharides. In our model, the ligand and binding site are represented in atomistic detail. The precise set of protein residues modelled atomistically is determined for a given ligand by the specific H-bonding and hydrophobic contacts between that ligand and the binding site. The protein model is not adaptive, i.e. the resolution of a given residue is {\it either atomistic or coarse-grained} and this does not change during the simulation. Solvent molecules, in contrast, may diffuse towards or away from the binding site, and are therefore modelled with {\it adaptive} resolution. Their atomistic or coarse-grained identity is determined by distance from the center of the binding site, such that the atomistic ligand and atomistic protein residues are solvated by a shell of atomistic water at least 1.2~nm thick.

\begin{figure*}[t]
\includegraphics[width=1.0\columnwidth,keepaspectratio]{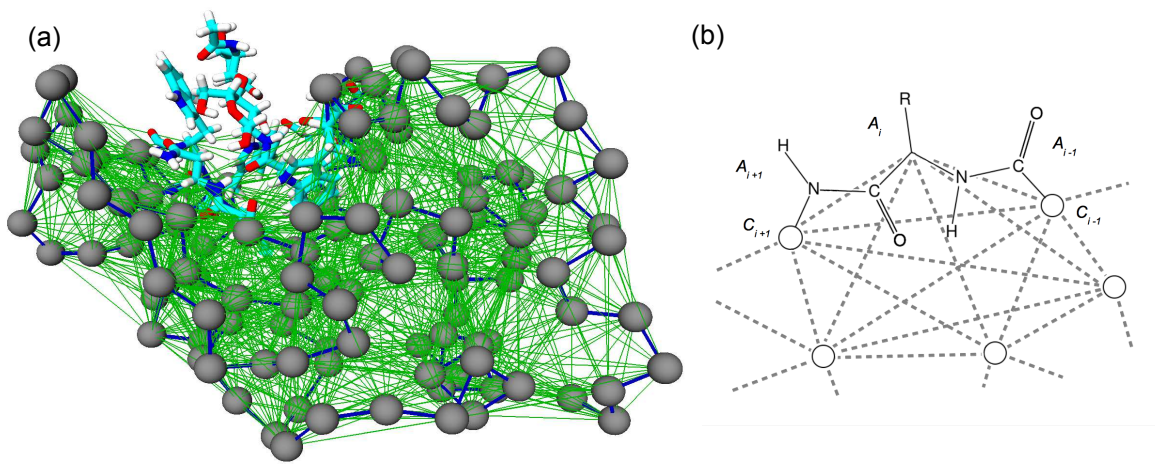}
\caption{(a) Visualisation of the multi-resolution protein model. The residues included in atomistic detail are shown in red, blue, cyan and white (O, N, C and H atoms). The grey spheres are ENM nodes, the stiff backbone springs are shown as dark blue lines and all other (weaker) springs are shown in green. (b) Bonded coupling between an atomistic protein residue and the coarse-grained protein model. Black lines and letters are bonds and atoms described using the atomistic forcefield, with 'R' representing any sidechain, circles are ENM nodes and dashed grey lines are ENM springs.}
\label{fig: model illustration}
\end{figure*}

We first describe bonded interactions within the protein, including parametrisation of a coarse-grained protein model and coupling between atomistic and coarse-grained protein. We then outline how non-bonded interactions in the protein and solvent are treated using the AdResS methodology. The model is summarised in Table~\ref{tab: model} and illustrated in Figure~\ref{fig: model illustration}.

\subsection{Bonded interactions in the multi-resolution protein model}
\label{subsec: bonded inter}

Proteins may be considered as having both local, high-frequency, small-amplitude fluctuations about conformational substates, and slower, more global transitions between them.\cite{Atilgan_BiophysJ_2002-proteins_glassy,McMahon_PNAS_1998-protdyn_gen} This is similar to the ``essential dynamics'' hypothesis pioneered by Berendsen.\cite{Berendsen_Proteins_1993-ess_dyn_prot}
In our minimalistic molecular modelling approach, only those local fluctuations which play a direct role in the biological function of interest are included on an atomistic level, i.e. in this case the ligand binding site. The set of protein degrees of freedom which are modelled atomistically does not change during the simulation, i.e. the protein is fixed-dual-resolution. 

The coarse-grained protein exists to ensure the correct structure and conformational fluctuations of the higher resolution binding site. To describe the coarse-grained protein we use an Elastic Network Model\cite{Tirion_PRL_1996-firstENM} in which each residue is mapped to a bead whose location corresponds to the C$_{\alpha}$ atom in the atomistic description. These beads are connected by harmonic springs. The potential energy is then given by

\begin{equation}
E = \sum_i\sum_j k_{ij}(r_{ij}-r_{ij}^0)^2h(r_c-r_{ij}^0)
\end{equation}

\noindent
with spring constants $k_{ij}$, equilibrium distances $r_{ij}^0$, a cutoff distance $r_c$, and where $i,j$ are nodes and $h(r) = 1$ if $r > 0$ and $h(r) = 0$ otherwise. This family of models has been shown to successfully capture global protein conformational fluctuations including low frequency modes.\cite{Grossfield_JCTC_2010-enm_comp,Orozco_JCTC_2010-enm_md} For the protein used here, lysozyme, this includes the widely studied hinge-bending motion.\cite{Wood_JBiolPhys_2014-lys_lig_rearrange}
Although the ENM is most often employed for Normal Mode Analysis, it can also be used in molecular dynamics simulations.\cite{Glenn_JCP_2015-ENM_MD} 

The ENM used here is parametrised such that it reproduces the conformational fluctuations of reference atomistic simulations, as quantified by the root mean square fluctuations (rmsf) of the C$_{\alpha}$ atoms and the NMR S$^2$ order parameters of the backbone NH bonds. Order parameters from $^{15}$N spin relaxation experiments are used for cross-validation and B-factors from X-ray crystallography structure determinations are used for comparison. (See the Supporting Material, which also contains some remarks on the relative reliability of different possible sources of reference data for parametrisation). While we primarily use atomistic reference simulations rather than experimental data here for parametrisation, this is only to facilitate the validation process. 

The coarse-grained ENM protein is then converted into a dual-resolution model as follows and as illustrated in Figure~\ref{fig: model illustration}. 
 For an N-residue protein with residues $A = \{A_i,i=1...N\}$ numbered along the protein backbone, a subset of residues $A'$ are designated as atomistic, and all backbone and sidechain atoms therein are modelled using a standard atomistic forcefield. The residues in $A'$ are not necessarily consecutive in the protein sequence. For a given atomistic residue $A_i \in A'$, if $A_{i-1} \notin A'$ then the backbone C and O atoms in $A_{i-1}$ are also modelled atomistically, and bonded to $C_{i-1}$ using standard atomistic forcefield parameters, with a corresponding reduction in the mass of the node $C_{i-1}$, and similarly for the backbone N and H atoms in $A_{i+1}$ if $A_{i+1} \notin A'$. The C$_{\alpha}$ atoms of the residues in $A'$ remain part of the coarse-grained network. For two C$_{\alpha}$ atoms with an atomistic bonded interaction, i.e. for consecutive atomistic residues along the protein backbone, the harmonic spring between the corresponding nodes is removed from the ENM. In short, atomistic residues are inserted directly into the ENM, with their C$_{\alpha}$ atoms replacing ENM nodes. This can be viewed as a step-wise change in resolution, as opposed to the gradual resolution change which takes place across the hybrid region in the solvent. Our procedure allows the coupling of coarse-grained and atomistic protein models without perturbing the structure and conformational fluctuations of either, as will be shown below.

\subsection{Non-bonded interactions in the multi-resolution system}
\label{subsec: nonbonded inter}

To treat the non-bonded interactions, we use the AdResS methodology,\cite{Kremer_JCP_2005-onthefly} which was developed to allow the simulation of multi-resolution fluids with free exchange of particles between an atomistic (AT) and a coarse-grained (CG) region. The theoretical basis\cite{Kremer_PhysRevLett_2012-thdforce,DelleSite_PhysRevX_2013-GCadres,DelleSite_JCP_2014-chempot_adres} and practical application\cite{Clementi_JPhysCondensMat_2007-adres_water,Kremer_JCP_2007-adres_macromolinsolvent} of this approach have been extensively explored, including its use in the simulation of systems containing fully atomistic biomolecules in solution.\cite{Praprotnik_JChemPhys_2014-adres_bio,Kremer_JCP_2015-adresprot,Kremer_JCTC_2012-triglyc} Here, we extend the methodology, describing how it applies in the case of a dual-resolution non-adaptive macromolecule solvated in adaptive resolution solvent.

A parameter $\lambda$ designates the resolution of each protein particle or water molecule. We set $\lambda = 1$ for atomistic protein or ligand atoms, and $\lambda = 0$ for coarse-grained protein particles, i.e. each node in the ENM. The $\lambda$ value for each water molecule depends on its location in space. A spherical region with radius $d_{at}$ is defined centred on a designated atom in the atomistic protein or ligand with coordinates $\mathbf{r}_{centre}$. Solvent molecules whose centre of mass $\mathbf{r}_{\alpha}$ satisfies $|\mathbf{r}_{centre}-\mathbf{r}_{\alpha}| < d_{at}$ have $\lambda = 1$ and are modelled at atomistic resolution. Solvent molecules with $|\mathbf{r}_{centre}-\mathbf{r}_{\alpha}| > (d_{at}+d_{hy})$ have $\lambda = 0$ and are modelled at a coarse-grained resolution, where each molecule is mapped to a single interaction site at its centre of mass. The quantity $d_{hy}$ is the thickness of a transition or hybrid (HY) region, where the value of $\lambda$ for solvent molecules varies monotonically from 1 to 0.\cite{Kremer_JCP_2005-onthefly} 

Nonbonded interactions forces are then calculated using the linear force interpolation

\begin{equation}
\mathbf{F}_{\alpha\beta} = \lambda(\mathbf{r}_{\alpha})\lambda(\mathbf{r}_{\beta})\mathbf{F}^{AT}_{\alpha\beta} + [1-\lambda(\mathbf{r}_{\alpha})\lambda(\mathbf{r}_{\beta})]\mathbf{F}^{CG}_{\alpha\beta}
\label{eq: force-adres}
\end{equation}

\noindent where

\begin{equation}
\mathbf{F}^{AT}_{\alpha\beta}=\sum_{i\in\alpha}\sum_{j\in\beta}\mathbf{F}^{AT}_{ij}
\end{equation}

\noindent and $\alpha,\beta$ are ENM beads or solvent molecule centres of mass, and $i,j$ are atoms in the atomistic protein or in the solvent molecules. 
In the current work, $\mathbf{F}^{AT}$ means the non-bonded contributions of any standard atomistic protein, ligand and water forcefields (here Amber99SB\cite{Simmerling_Prot_2006-ff99sb}, GLYCAM\cite{Woods_JCompChem_2008-glycam06} and SPCE/E\cite{Berendsen_jphyschem_1987-SPCE}), and $\mathbf{F}^{CG}$ means any standard coarse-grained model for water-water interactions (here a potential derived via Iterative Boltzmann Inversion\cite{Soper_ChemPhys_1996-IBI,MullerPlathe_JComputChem_2003-IBI}) and an excluded-volume interaction between protein and water (see Supporting Material for further details on all potentials). This scheme leads to the following interactions. Non-bonded interactions within the atomistic protein, between two atomistic water molecules, and between atomistic protein and water simplify to $\mathbf{F}^{AT}_{ij}$.
There are no non-bonded interactions between the coarse-grained and atomistic protein, nor within the coarse-grained protein, these interactions being modelled entirely by the ENM.
Interactions between two coarse-grained water molecules simplify to $\mathbf{F}^{CG}_{\alpha\beta}$. 
Interactions within the solvent across the resolution boundaries are treated using the AdResS interpolation (Eq.~\ref{eq: force-adres}).
Interactions between the atomistic protein and hybrid or coarse-grained water also use the interpolation, however in practice $d_{at}$ should be chosen to be large enough so these do not occur. 
Finally, all water molecules regardless of their resolution have an excluded volume interaction with the coarse-grained protein nodes.

This force-interpolation scheme is inherently non-conservative\cite{DelleSite_PhysRevX_2013-GCadres,DelleSite_JCP_2014-chempot_adres} and a local thermostat must be employed in order to remove the excess heat produced in the hybrid region. An alternative, energy-conserving adaptive resolution scheme exists in which energies rather than forces are interpolated,\cite{Donadio_PRL_2013-hadres_molliq} however in this case momentum is no longer strictly conserved. In other words, in the adaptive resolution solvent, one has to choose either energy or momentum conservation. This is a well known issue,\cite{DelleSite_PhysRevE_2007-forceenergypb} and does not prevent practical applications. We clarify that this question of choosing between a force-based or energy-based scheme does not arise in the case of the model for bonded interactions in the protein, which is non-adaptive, and is both energy-conserving (that is, a potential energy is well defined) and momentum-conserving.

The AdResS scheme has been demonstrated to yield an atomistic region whose structural and dynamical properties are identical to those of an equivalent region in a much larger fully atomistic system.\cite{Praprotnik_JChemPhys_2014-adres_bio,Kremer_JCP_2015-adresprot,Jayaraman_JPCB_2016-adresPEG,Kremer_JCP_2005-onthefly,DelleSite_JCP_2008-diffusive_water,DelleSite_PhysRevX_2013-GCadres} Moreover, this is independent of the quality of the solvent coarse-grained model chosen. In fact, we recently demonstrated that two models as different as atomistic water and a gas of non-interacting particles can be coupled via the AdResS methodology without perturbing the properties of the atomistic water.\cite{Kremer_EPJ_2015-idealgas} One simply needs a reservoir of coarse-grained particles such that particles are correctly supplied to and accepted from the atomistic region.

\section{Results and discussion}
\label{sec: results}

\begin{figure}[t]
  \centering
\includegraphics[clip,width=0.50\columnwidth,keepaspectratio]{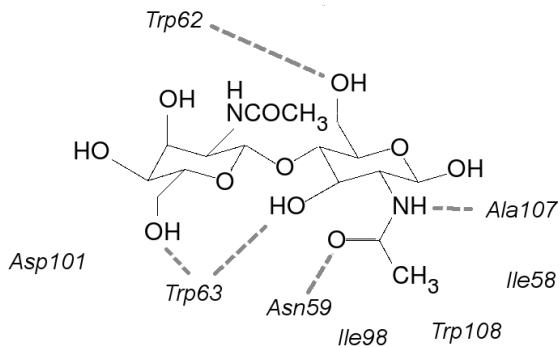}
  \caption{Atomistic residues and inhibitor in the binding site of the protein. Dashed grey lines are hydrogen bonds.}
\label{fig: act site}
\end{figure}

We performed fully atomistic and multi-resolution molecular dynamics simulations of HEWL, with and without the ligand di-N-acetylchitotriose, an inhibitor. This enzyme hydrolyses glycosidic bonds in polysaccharides. The binding site is in a cleft between two lobes and has both chemical and steric specificity, both of which our model will reproduce. In the multi-resolution enzyme, the residues modelled on an atomistic level are those that form stable H-bonds to the ligand (Asn-59, Trp-62, Trp-63 and Ala-107) and the four nearest neighbours in terms of the distance of their centre of mass from the ligand (Ile-58, Ile-98, Asp-101, Trp-108, see Supporting Material). This is illustrated in Figure~\ref{fig: act site}.

For the study of processes such as inhibitor binding or enzyme catalysis using a multi-resolution model, the ligand must experience an environment as close as possible to the true environment. Here, we examine the properties of the ligand and the binding site in the multi-resolution model, comparing them to those determined experimentally or via fully atomistic simulations. In this context, our current understanding of the biological process of interest must be borne in mind. For example, ligand binding may be understood via the opposing induced-fit or population-shift models.\cite{Takada_PNAS_2008-inducefit_popselect} Similarly, for enzyme catalysis, opposing hypotheses posit either stabilisation of the transition state by a ``pre-organised'' environment, or coupling of certain vibrational modes in the enzyme to the enzymatic reaction coordinate.\cite{Warshel_Proteins_2010-dawn21st,Truhlar_science_2004-howenzwork}
Such open questions can be seen as a challenge to be overcome in the construction of a multi-resolution model, or rather as an opportunity for multi-resolution models to contribute to the debate.
As a first step, the multi-resolution model must reproduce the relevant enzyme properties in all cases. We first discuss the global properties of the protein, before turning to the properties of the binding site.

\subsection{Global protein structure and fluctuations}

\begin{figure}[t]
  \centering
\includegraphics[clip,width=0.50\columnwidth,keepaspectratio]{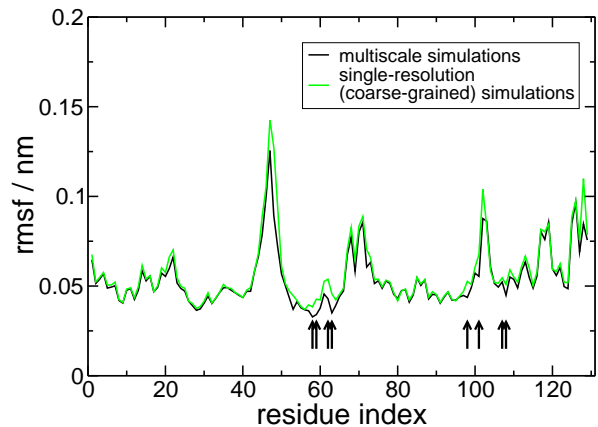}
  \caption{Rmsf of beads in the single-resolution ENM in single-resolution solvent vs rmsf of ENM beads and protein C$_{\alpha}$ atoms in the multi-resolution system. The black arrows mark the positions of the atomistic residues in the multi-resolution case.}
\label{fig: rmsf enm vs enmaa}
\end{figure}

The protein's global properties determine the local properties of the binding site. In the ENM, the global structure of the protein is assured by construction. As an aside, although an ENM has only one equilibrium structure, proteins with two or more distinct and well-defined conformational states on a global level could be represented using a double or multi-well ENM.\cite{Voth_BiophysJ_2007-doublewell_netmod} The global conformational fluctuations of the coarse-grained model, controlled by $k_{ij}$, were parametrised in a single-resolution, fully coarse-grained ENM in the first step of model-building, and these parameters were then used in the multi-resolution model. Figure~\ref{fig: rmsf enm vs enmaa} shows the rmsf of the ENM beads and C$_{\alpha}$ atoms in the multi-resolution model versus the rmsf of the ENM beads in the single-resolution ENM. The global fluctuations of the protein, as parametrised in the purely coarse-grained model, are clearly not perturbed by the insertion of atomistic detail. We note that the rmsf values of the C$_{\alpha}$ atoms in the multi-resolution model, marked by black arrows, cannot be compared directly to the rmsf values of the corresponding ENM nodes in the single-resolution model, as these represent the movement of an entire residue. 

\subsection{Properties of the binding site in the ligand-free system}

We now turn to key properties of the binding site, in each case comparing multi-resolution and fully atomistic simulations. 

\begin{figure}[t]
  \centering
\includegraphics[clip,width=0.50\columnwidth,keepaspectratio]{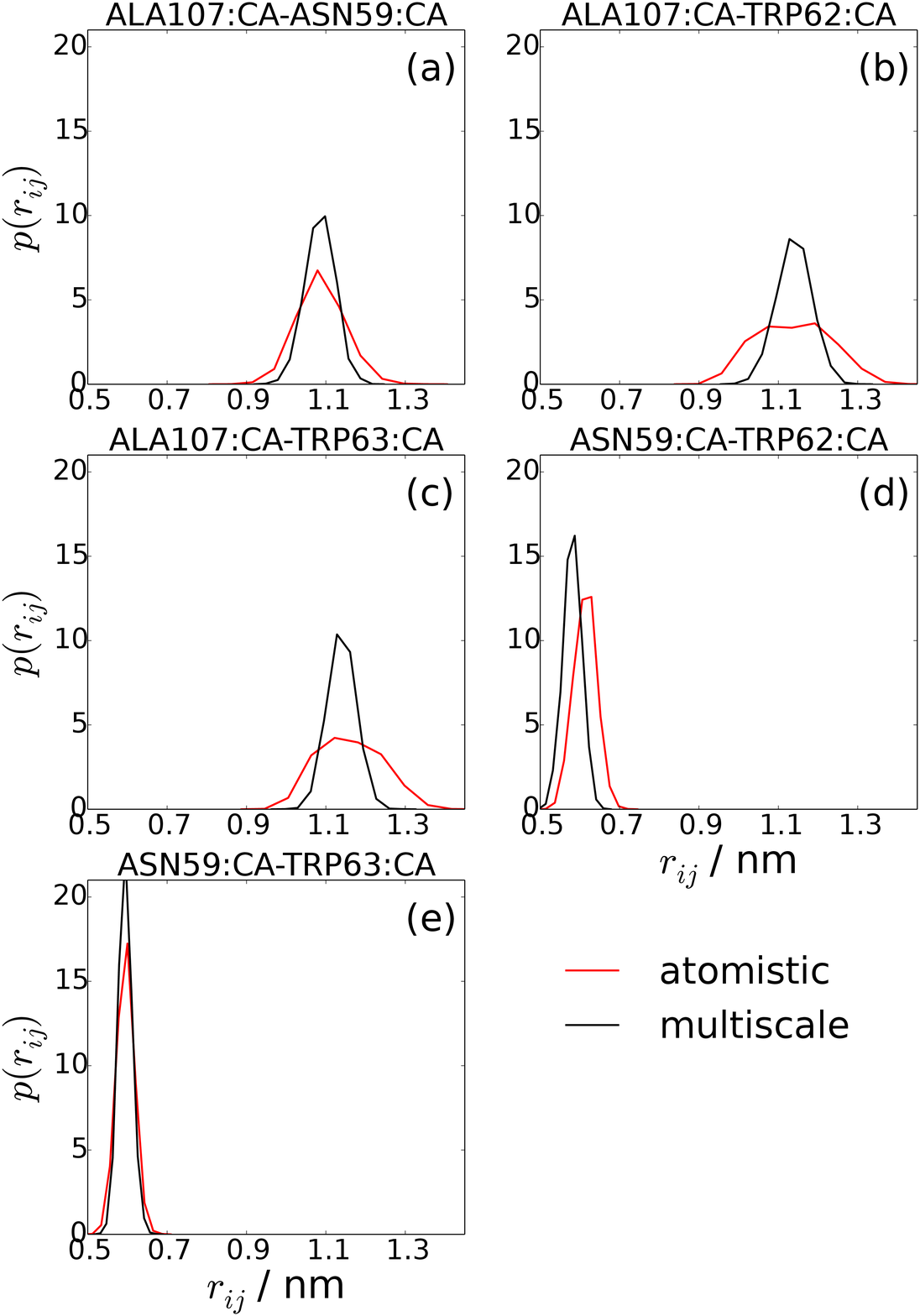}
\caption{Probability density distributions for key protein-protein distances across the binding site in the ligand-free system, multi-resolution versus fully atomistic simulations.}
\label{fig: actsite fluct dist}
\end{figure}

Starting with the ligand-free system, we study the structure and conformational fluctuations of the binding site via the distributions of distances between the four C$_{\alpha}$ atoms in the core H-bonding residues (Asn-59, Trp-62, Trp-63 and Ala-107). 
 These are shown in Figure~\ref{fig: actsite fluct dist}.

Residues Asn-59, Trp-62 and Trp-63 are in the enzyme's $\beta$-lobe, while Ala-107 is in the $\alpha$-lobe. The first three distances shown (Ala107:CA-Asn59:CA, Ala107:CA-Trp62:CA and Ala107:CA-Trp63:CA, Figure~\ref{fig: actsite fluct dist}(a-c)) cross the binding cleft. The agreement between atomistic and multi-resolution structures, as measured by the distributions' maxima, is very good; however, the distributions in the multi-resolution system are considerably narrower. Here, we see a result of the harmonic approximation inherent in the ENM. The protein fluctuations as described by the fully atomistic forcefield have a much greater variance, and while the distributions examined here remain reasonably symmetric and monomodal, this will not necessarily be the case for fluctuations across the active site of all proteins. The remaining two distances shown (Figure~\ref{fig: actsite fluct dist}(d,e)) are within the $\beta$-lobe. Again, the agreement between atomistic and multi-resolution simulations is good, within the limits of the harmonic ENM model.
The ENM used here has only two spring constants, $k_b$ for $i...i+1$ interactions along the backbone and $k_{nb}$ for all other interactions. The agreement between multi-resolution and atomistic fluctuations could of course be improved by the use of a more complex ENM model, or at least by tuning specific spring constants, for example for $i...i+2$ and $i...i+3$ interactions along the backbone. However such detailed information on the local conformational fluctuations is not available experimentally. Instead of overfitting our model to perhaps unreliable atomistic simulations, we prefer to retain simplicity, and therefore applicability to a wide range of systems. This simple but widely used ENM is well known to successfully model global protein fluctuations.\cite{Grossfield_Proteins_2011-enm_md} In the context of our multi-resolution model, it supplies the global structure and fluctuations that determine the local properties of the binding site, allowing modelling of the phenomenon of molecular recognition, as shown below.

A unique feature of our methodology is its incorporation of the AdResS treatment of the solvent, allowing us to solvate the atomistic binding site in a small sphere of atomistic water, which is in turn immersed in a less expensive coarse-grained solvent, but which behaves as though it were in a fully atomistic solvent. The sphere is centred on an atom of the protein or ligand (see Supporting Material) and its radius $d_{at}$ is 2.4~nm, such that there is always at least 1.2~nm between any atomistic protein or ligand atom and the atomistic/hybrid boundary. A large portion of this sphere is occupied by the excluded volume of the protein, and it contains on average 1450 fully atomistic water molecules. This number rises to 3250 if we include all water molecules with $\lambda > 0.5$, i.e. with at least 50 \% atomistic character. To confirm that this is sufficient to reproduce the behaviour of a fully atomistic simulation, we now examine the properties of water in the system and the hydration of the ligand-free binding site.

We first calculate the water density across the simulation box. This is done by dividing the box into a three-dimensional grid of subcells, assigning subcells to the AT, HY and CG region at each timestep, excluding those subcells which overlap two regions or contain protein excluded volume, and then averaging the recorded values in time for each region. The bulk water density is 0.995 $\pm$ 0.002, 1.001 $\pm$ 0.001 and 0.996 $\pm$ 0.001 g cm$^{-3}$ in the AT, HY and CG regions respectively. This is in excellent agreement with the reference value of 1.000 $\pm$ 0.001 g cm$^{-3}$ from the fully atomistic simulations of the same system.

\begin{figure}[t]
  \centering
\includegraphics[clip,width=0.35\columnwidth,keepaspectratio]{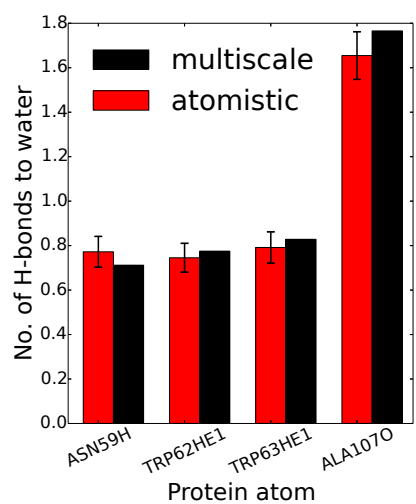}
\caption{Number of H-bonds between water and the key H-bond acceptors and donors in the binding site, in the ligand-free system, multi-resolution versus fully atomistic simulations.}
\label{fig: hbcount}
\end{figure}

We then examine the hydration of the binding site. We identify the four atoms which form stable H-bonds with the inhibitor when it is present: Asn-59:H, Trp-62:HE1, Trp-63:HE1 and Ala-107:O. In the inhibitor-free system, we then calculate the average number of hydrogen bonds to water formed at those key sites, i.e. these are H-bonds to the water molecules which are displaced by ligand binding. Figure~\ref{fig: hbcount} shows the comparison between fully atomistic and multi-resolution simulations. The agreement is excellent for the three donor and one acceptor sites. 
We note that the correct modelling of the binding site hydration is made possible by the AdResS treatment of the solvent, which ensures that the density and other properties of water are correctly modelled even with only a finite sphere of atomistic water around the binding site. While we do not perform an exhaustive study of water properties here, it has been shown elsewhere that the structure and dynamics of water molecules hydrating protein surfaces can be well reproduced with the AdResS setup.\cite{Kremer_JCP_2015-adresprot}

\subsection{Properties of the binding site in the ligand-bound system}

\begin{figure}[t]
  \centering
\includegraphics[clip,width=0.45\columnwidth,keepaspectratio]{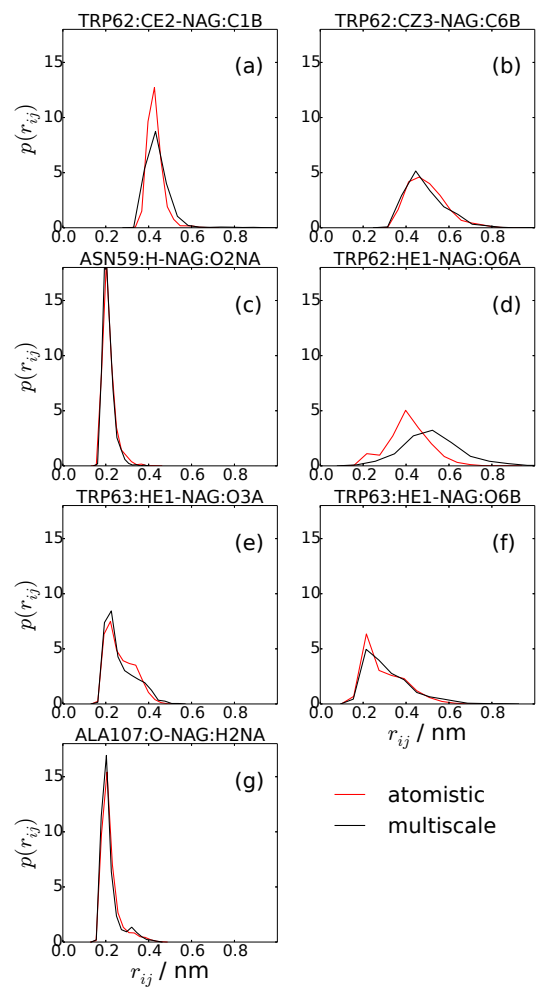}
\caption{Probability density distributions for the key distances in the hydrophobic contacts (a,b) and the H-bonds (c-g) between the protein and the ligand, multi-resolution versus fully atomistic simulations.}
\label{fig: hbonding dist}
\end{figure}

We now turn to the interactions between the inhibitor and the protein. Figure~\ref{fig: hbonding dist} shows the probability density distributions for the key distances which quantify contacts between the binding site and the inhibitor. This includes two distances measuring the hydrophobic contact between the Trp-62 aromatic sidechain and the ligand (Figure~\ref{fig: hbonding dist}(a,b)) and the heavy atom-hydrogen distances in five H-bonds (to Asn-59:H, Trp-62:HE1, Trp-63:HE1 twice, and Ala-107:O, Figure~\ref{fig: hbonding dist}(c-g)). The inhibitor clearly remains stably bound to the dual-resolution protein over the entire 14-ns simulation, forming inhibitor-protein contacts exactly as in the fully atomistic model. For the one case where the comparison is good but not perfect (Trp-62:HE1--NAG:O6A, Figure~\ref{fig: hbonding dist}(d)), the H-bond is weak, and is only formed a small portion of the time. While the H-bond is broken, this distance samples a large conformational space, making it more difficult to reproduce.

\begin{figure}[t]
  \centering
\includegraphics[clip,width=0.9\columnwidth,keepaspectratio]{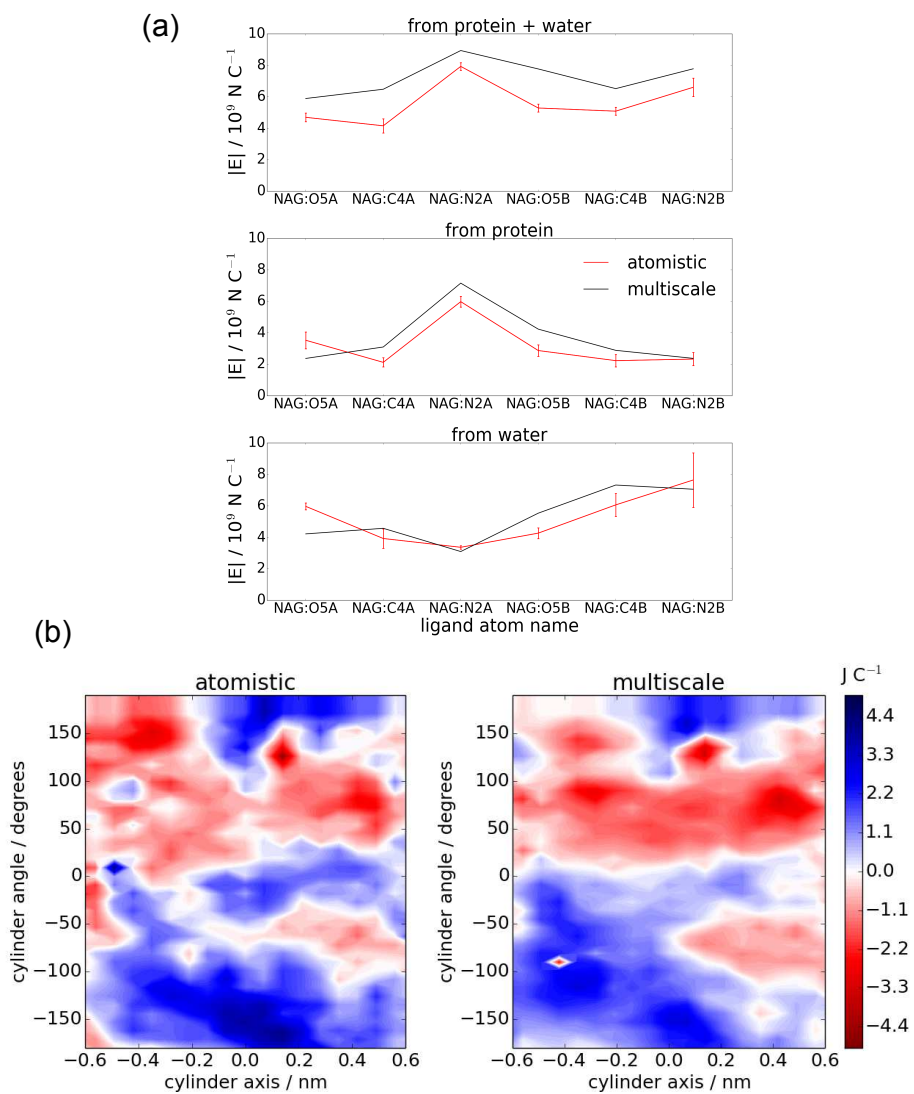}
\caption{(a) Electric field on selected atoms in the ligand, from the protein and from water, multi-resolution versus fully atomistic simulations. (b) Electrostatic potential felt by the ligand in the binding site due to protein and solvent charges, on the surface of a cylinder of radius 0.3~nm enclosing the ligand. The cylinder is centered on the ligand center of mass and its axis runs along the ligand's longest dimension, such that the cylinder's size, shape and orientation match that of the ligand.}
\label{fig: elecfield onNAG}
\end{figure}

According to the electrostatic stabilisation hypothesis,\cite{Truhlar_science_2004-howenzwork} the electric field of the protein plays a key role in enzyme catalysis by stabilising the transition state of the reaction. Studies from the field of non-aqueous enzymology suggest that the polarity of the aqueous solvent also plays a role. 
In Figure~\ref{fig: elecfield onNAG}, we show the magnitude of the electric field from the protein and from water on key atoms in the ligand, in the fully atomistic and multi-resolution simulations. In the same figure we also show the electrostatic potential in the binding site, calculated on the surface of a cylinder enclosing the ligand.\cite{axes_note} The potential at each point in the plane is calculated as a summation over atomistic charges, in order to allow a comparison between the particle-based fully atomistic and multi-resolution models. The electrostatic potential shown is that due to atomistic protein and solvent charges, excluding charges of the ligand, i.e. that felt by the ligand due to its environment. Again, the agreement is good, completing the demonstration that the environment felt by the ligand in the multi-resolution model matches that in a fully atomistic model. Were the multi-resolution model used to study enzymatic catalysis, the transition-state stabilisation effect would be satisfactorily included.

The analysis outlined above demonstrates that our model can accurately capture the interaction of a ligand with the protein's binding site. To do so, a sufficient but still very small number of residues must be included in atomistic detail. The eight atomistic residues used here contain only 8.5\% of the total atomistic degrees of freedom in the protein. The precise details of this will vary from enzyme to enzyme and will form the subject of a future work.

\section{Conclusion}
\label{sec: conclude}

We have developed a multi-resolution methodology in which arbitrary degrees of freedom in a protein/water system can be included in atomistic detail. The multi-resolution model captures the global properties of the protein via a highly coarse-grained model, while still allowing for chemical detail in the ligand binding site. This highly minimalistic model is enough to model stable ligand binding with the inclusion on an atomistic level of only 8.5\% of the enzyme's total atomistic degrees of freedom. For larger proteins this proportion is expected to be even lower.

Our methodology is transferable to any atomistic protein and solvent force field. As discussed earlier, the choice of the coarse-grained solvent force field has no effect on the properties of the atomistic region. Regarding the coarse-grained protein model, the ENM used here could be replaced by a very different ansatz, such as a G\={o}-like model or multi-bead protein model,\cite{Tozzini_CurrOp_2005-CGmodels_prot} in order to capture different protein features, for example large-scale conformational change or side-chain flexibility. Coarse-graining always represents a compromise, and the choice of coarse-grained protein model will be dictated by the biophysical or biochemical processes being studied in any given application.

Our approach to the coarse-grained protein allows a parametrisation directly on experiment, instead of using as a reference a possibly problematic atomistic forcefield. Although atomistic protein force fields form an essential part of the biomolecular simulation toolkit, they also have well-known weaknesses, and can deviate from the experimentally observed behaviour with increasing trajectory length, due to the accumulation of errors in the parametrization of the force field.\cite{Zagrovic_PLOSCompBiol_2014-atff_accuracy,Shaw_CurrOpStructBiol_2014-biomolffields,Schulten_BiophysJ_2009-ffprob} In our approach, the use of the atomistic force field is limited to the binding site region. Moreover, this allows the simulation of systems for which the structure is known in high resolution only in some parts of the system, as well as systems that are too large for a fully atomistic modelling.

This opens the way to a range of applications. In structure-based drug design, for example, docking or molecular recognition studies must use simulation methods which are as efficient as possible, due to the large number of lead compounds which must be screened.\cite{Higo_Proteins_2007-flexdocking} The receptor is usually treated as a rigid body for reasons of computational cost, however it has been shown that receptor flexibility and thermal motions play an important role in molecular recognition.\cite{Higo_Proteins_2007-flexdocking,Carlson_JACS_2011-hotspot_flex} The model presented here could be used to combine both flexibility and computational efficiency.

Beyond computational efficiency, a model which allows the inclusion and exclusion of arbitrary degrees of freedom while still correctly modelling structure, dynamics and thermodynamics can be used to pinpoint and quantify the contribution of any given set of degrees of freedom to biophysical or biochemical processes in the system, for example free energies of binding or enzymatic reaction rate constants. Enzymatic catalysis could be studied using the current purely classical methodology via the Empirical Valence Bond method,\cite{Warshel_Faraday_2010-evb_bio} without needing the inclusion of a quantum/classical coupling in addition to the classical atomistic/coarse-grained coupling. 

Future refinement to the multi-resolution model presented here will include a coarse-grained model capable of capturing anharmonic fluctuations, and inclusion of counterions in the AdResS treatment of the solvent.

\section{Author contributions}

A.C.F., R.P. and K.K. designed research; A.C.F. performed research; A.C.F., R.P. and K.K. wrote the manuscript.

\section{Supporting Material}

Additional supplemental information including three figures, parametrisation of the coarse-grained models, and simulation details are available online at \url{http://onlinelibrary.wiley.com/wol1/doi/10.1002/prot.25173/full}.

\section{Acknowledgements}

K.K. and A.C.F. acknowledge research funding through the European Research Council under the European Union's Seventh Framework Programme (FP7/2007-2013) / ERC grant agreement no. 340906-MOLPROCOMP. We are grateful to Torsten Stuehn for assistance with the ESPResSo++ package and to Debashish Mukherji and Tristan Bereau for a critical reading of the manuscript.

\bibliographystyle{rsc} 

\begin{mcitethebibliography}{58}
\bibitem[{For the electrostatic potential calculations}()]{axes_note}
{For the electrostatic potential calculations}, a coordinate system was defined
  in the frame of reference of the ligand. Its origin was at the ligand centre
  of mass, and its axes were the first three principal components of the
  covariance matrix of a set of four ligand heavy atoms in the sugar
  rings.\relax
\mciteBstWouldAddEndPunctfalse
\mciteSetBstMidEndSepPunct{\mcitedefaultmidpunct}
{}{\mcitedefaultseppunct}\relax
\EndOfBibitem
\end{mcitethebibliography}
\providecommand*{\mcitethebibliography}{\thebibliography}
\csname @ifundefined\endcsname{endmcitethebibliography}
{\let\endmcitethebibliography\endthebibliography}{}

\begin{table*}[h]
    \begin{tabular}{|l||l|l|l|l|l|}
\hline
          & AT protein & ENM protein & AT water & HY water & CG water       \\
\hline
\hline
    AT protein      & $F_{AT}$   & ~      & ~       & ~    & ~         \\
\hline
    ENM protein     & $F_{ENM}$   & $F_{ENM}$  & ~     & ~   & ~     \\
\hline
    AT water        & $F_{AT}$    & $F_{WCA}$ &     $F_{AT}$   & ~    & ~     \\
\hline
    HY water        & $\lambda F_{AT}+$  & $F_{WCA}$  & $\lambda F_{AT}+                  $ & $\lambda F_{AT}+                  $ & ~              \\
                    & $(1-\lambda)F_{WCA}$ [1] & & $(1-\lambda)F_{IBI}$ & $               (1-\lambda)F_{IBI}$ & ~              \\
\hline
    CG water        & $F_{WCA}$ [1]        & $F_{WCA}$ & - [2]                                                        & $F_{IBI}$                                               & $F_{IBI}$ \\
\hline
\multicolumn{6}{|c|}{ [1] not used if $d_{at}$ is large enough, [2] because $d_{hy} \geq$ non-bonded cutoff} \\
\hline
    \end{tabular}
\caption{Summary of interactions in the multi-resolution model. $F_{ENM}$ = Elastic Network Model; $F_{AT}$ = atomistic AMBER+GLYCAM forcefield and SPC/E water; $F_{CG}$ = $F_{WCA}$, $F_{IBI}$, where $F_{WCA}$ = WCA (excluded volume) interaction and $F_{IBI}$ = potential from IBI coarse-graining of atomistic (SPC/E) water; AT protein: $\lambda$ = 1, ENM protein: $\lambda$ = 0, water: $0 < \lambda < 1$}
\label{tab: model}
\end{table*}

\end{document}